\documentclass{article}

\usepackage[english]{babel}
\usepackage[letterpaper,top=2cm,bottom=2cm,left=3cm,right=3cm,marginparwidth=1.75cm]{geometry}

\usepackage{amsmath}
\usepackage{graphicx}
\usepackage[colorlinks=true, allcolors=blue]{hyperref}
\usepackage{booktabs}
\usepackage{tablefootnote}
\usepackage{multirow}
\usepackage{subcaption}

\title{Confidential Computing on NVIDIA Hopper GPUs: A Performance Benchmark Study}
\author{
  Jianwei Zhu,
  Hang Yin,
  Peng Deng$^\dagger$,
  Aline Almeida$^\ddagger$,
  Shunfan Zhou \\
  Phala Network, $^\dagger$Fudan University, $^\ddagger$io.net \\
  \texttt{\{jianweiz, hangyin, shelvenzhou\}@phala.network}, \\
  \texttt{$^\dagger$pdeng21@m.fudan.edu.cn},
  \texttt{$^\ddagger$aline@io.net}
}

\begin{document}
\maketitle

\begin{abstract}
    This report evaluates the performance impact of enabling Trusted Execution Environments (TEE) on NVIDIA Hopper GPUs for large language model (LLM) inference tasks. We benchmark the overhead introduced by TEE mode across various LLMs and token lengths, with a particular focus on the bottleneck caused by CPU-GPU data transfers via PCIe. Our results indicate that while there is minimal computational overhead within the GPU, the overall performance penalty is primarily attributable to data transfer. For the majority of typical LLM queries, the overhead remains below 7\%, with larger models and longer sequences experiencing nearly zero overhead.
\end{abstract}

\section*{Acknowledgments}

We would like to express our gratitude to the io.net~\cite{ionet} and IOG Foundation~\cite{iog} for their generous grant, which made this research possible. We also extend our thanks to Engage Stack~\cite{engagestack}, the cloud service provider, for providing the necessary hardware and technical support.

\section{Introduction}

Trusted Execution Environments (TEEs) are increasingly important in machine learning and AI due to growing security requirements in both enterprise and decentralized applications~\cite{sabt2015trusted, matetic2018delegatee, ayoade2018decentralized}. The introduction of TEE-enabled GPUs, such as the NVIDIA H100 and H200, adds an extra layer of protection for sensitive data but may impact performance. Understanding these trade-offs, particularly for large-scale machine learning tasks, is crucial for adopting TEE in high-performance AI applications~\cite{yudha2022lite, wang2024confidential}.

This report quantifies the performance overhead of enabling TEE on the NVIDIA Hopper architecture GPUs during LLM inference tasks, identifying where the overhead arises and under what conditions it can be minimized.

\section{Background}

\subsection{Trusted Execution Environment}

A TEE is a hardware-based security feature that isolates computations, preventing unauthorized access and tampering, even from the operating system or the physical hardware owner. As the core technology enabling Confidential Computing, TEEs create secure enclaves where sensitive data and code are processed with encryption, ensuring confidentiality and integrity even if the broader system is compromised~\cite{sabt2015trusted}. Traditionally implemented in CPUs, TEE technology was extended to GPUs by NVIDIA in 2023, enabling tamper-proof and confidentiality-preserving computation inside the GPU with minimal performance penalty~\cite{dhanuskodi2023creating}.

\subsection{NVIDIA Hopper Architecture}

The NVIDIA Hopper architecture marks a significant milestone as the first GPU family to support TEE~\cite{nvidiacc}. In TEE mode, the GPU operates in an isolated and secure environment where data transfers between the CPU and GPU are encrypted. This is achieved through ``bounce buffers'', which protect all inputs and outputs during transit between the CPU’s encrypted memory and the GPU’s internal memory~\cite{dhanuskodi2023creating}.

To maintain end-to-end security, the GPU works in conjunction with CPU TEEs, such as Intel’s TDX~\cite{TDX} or AMD’s SEV-SNP~\cite{SEV-SNP, sev2020strengthening}, securing communication channels between the GPU driver and interacting software. This setup prevents unauthorized access and ensures data integrity throughout the process.

The Hopper series also implement remote attestation to verify the GPU’s identity and the authenticity of its firmware. Additionally, Secure Boot ensures that only authenticated firmware is executed during the GPU’s boot process, further strengthening security.

\subsection{Performance Impact}

Enabling TEE on the NVIDIA Hopper GPU introduces performance overheads primarily due to additional encryption and decryption during secure data transfer~\cite{mohan2024securing}. While the GPU’s internal computation remains unaffected, the main bottleneck lies in the CPU-GPU I/O, particularly when data is exchanged via PCIe. This impact varies with the size of the data transfer. The following sections present experimental results quantifying these effects across various use cases.

With the TEE-enabled NVIDIA Hopper GPU, it becomes crucial to quantify performance trade-offs during practical use cases. In the next section, we outline the methodology used to assess the performance impact of TEE during LLM inference tasks.

\section{Methodology}

To evaluate the performance overhead, we conducted experiments comparing inference throughput and latency with TEE mode enabled and disabled, under different models, input and output lengths, and batch size setups. Our primary focus was to reveal the performance penalty in real-world large language model (LLM) inference tasks.

\subsection{Metrics}

The primary metrics were evaluated following typical evaluation frameworks \cite{agrawal2024metron}:

\begin{itemize}
    \item \textbf{TTFT (Time To First Token):} The time from request arrival to the generation of the first output token. It includes scheduling delay and prompt processing. Lower TTFT is essential for real-time applications, while higher TTFT is tolerable in batch processing.

    \item \textbf{ITL (Inter-Token Latency):} The time between generating each token during decoding. This directly affects the perceived model speed. A rate of around 6 tokens per second is necessary for a smooth user experience, assuming an average reading speed.

    \item \textbf{TPS (Tokens per Second):} The average rate of token generation during decoding. It is calculated as the number of tokens generated divided by the total decoding time.

    \item \textbf{Latency:} The total execution time per request, including scheduling, prompt processing, and token generation. Lower normalized latency improves system throughput, especially under high query loads.

    \item \textbf{QPS (Queries per Second):} The maximum load a system can handle while meeting latency targets. Higher QPS reduces serving costs and is a key measure of system capacity.
\end{itemize}

\subsection{Test Scenarios}

Experiments were structured to explore the impact of TEE mode under diverse conditions:
\begin{itemize}
    \item \textbf{TEE mode ON vs. TEE mode OFF}: Tests were performed with TEE mode alternately enabled and disabled on the H-series GPUs, allowing for a direct comparison of performance.
    \item \textbf{Sequence Lengths}: Various token lengths were tested by sampling the ShareGPT Dataset \cite{sharegpt} to simulate different LLM inference tasks.
    \item \textbf{Batch Size}: Both fixed batch sizes (1, 4, and 16) and dynamic batch sizes determined by vLLM \cite{kwon2023efficient} were tested to simulate the performance for serving real-time requests and batch requests.
\end{itemize}

\subsection{Experimental Setup}

\subsubsection{Infrastructure}

The experiments were set up with the following hardwares, respectively.

\begin{table}[htbp]
    \centering
    \begin{tabular}{llcc}
        \toprule
        \textbf{Component}                 & \textbf{Specification} & \textbf{Setup 1} & \textbf{Setup 2}         \\
        \midrule
        \multirow{3}{*}{\textbf{GPU}}      & Model                  & NVIDIA H100 NVL  & NVIDIA H200 NVL          \\
                                           & Memory                 & 94 GB            & 141 GB                   \\
                                           & Bandwidth              & 3.9 TB/s         & 4.8 TB/s                 \\
        \midrule
        \multirow{3}{*}{\textbf{CPU}}      & Model                  & AMD EPYC 9V84    & INTEL XEON PLATINUM 8558 \\
                                           & Cores                  & 96               & 48                       \\
                                           & TEE Technology         & SEV-SNP          & TDX                      \\
        \midrule
        \textbf{Memory}                    & Total Memory           & 314 GB           & 128 GB                   \\
        \midrule
        \multirow{3}{*}{\textbf{Software}} & CUDA Version           & 12.5             & 12.5                     \\
                                           & Driver Version         & 555.42.06        & 555.42.06                \\
                                           & Kernel Driver Version  & 550.90.07        & 550.90.07                \\
        \bottomrule
    \end{tabular}
    \caption{Hardware Setup}
    \label{tab:hardware_setup}
\end{table}

\subsubsection{Application}

The experiments utilized the benchmark suite of \texttt{vLLM v0.5.4 (rev: 4db5176)} \cite{kwon2023efficient}.

\subsubsection{Models}

Three LLMs were used for inference:

\begin{itemize}
    \item \textbf{Meta-Llama-3.1-8B-Instruct}
    \item \textbf{Phi-3-14B-128k-Instruct}
    \item \textbf{Meta-Llama-3.1-70B-Instruct} with 4-bit bitsandbytes quantization to fit into a single Hopper GPU
\end{itemize}

\section{Results}

\noindent\textbf{Conclusion 1: The average overhead is less than 9\%.} We quantified the overhead by measuring the throughput with TEE mode enabled versus disabled, across varying input sizes and model configurations, as shown in Table \ref{tab:throughput_tee_on_ff}.

\begin{table}[htbp]
    \centering
    \begin{tabular}{llccc|ccc}
        \toprule
        \textbf{GPU}                   & \textbf{Model}         & \multicolumn{3}{c}{\textbf{TPS (tokens/s)}} & \multicolumn{3}{c}{\textbf{QPS (req/s)}}                                                                                                                                                                                                \\
                                       &                        & \textbf{TEE-on}                             & \textbf{TEE-off}                         & \textbf{Overhead}                                                          & \textbf{TEE-on} & \textbf{TEE-off} & \textbf{Overhead}                                                          \\
        \midrule
        \multirow{3}{*}{\textbf{H100}} & \textbf{LLama-3.1-8B}  & 123.2985                                    & 132.3618                                 & 6.85\%                                                                     & 18.2141         & 18.8208          & 3.22\%                                                                     \\
                                       & \textbf{Phi3-14B-128k} & 66.5845                                     & 69.7787                                  & 4.58\%                                                                     & 7.1760          & 7.3456           & 2.31\%                                                                     \\
                                       & \textbf{Llama-3.1-70B} & 2.4822                                      & 2.4789                                   & -0.13\%\tablefootnote{The overhead is negative due to the precision loss.} & 0.8325          & 0.8295           & -0.36\%\tablefootnote{The overhead is negative due to the precision loss.} \\
        \midrule
        \multirow{3}{*}{\textbf{H200}} & \textbf{LLama-3.1-8B}  & 121.0412                                    & 132.7830                                 & 8.84\%                                                                     & 29.5973         & 32.0134          & 7.55\%                                                                     \\
                                       & \textbf{Phi3-14B-128k} & 68.4287                                     & 72.9825                                  & 6.24\%                                                                     & 12.8294         & 13.8558          & 7.41\%                                                                     \\
                                       & \textbf{Llama-3.1-70B} & 4.0797                                      & 4.1753                                   & 2.29\%                                                                     & 2.1874          & 2.2011           & 0.63\%                                                                     \\
        \bottomrule
    \end{tabular}
    \caption{Performance comparison of TEE-on and TEE-off modes for various models in terms of TPS (tokens per second) and QPS (queries per second).}
    \label{tab:throughput_tee_on_ff}
\end{table}

The throughput is measured in two ways: the average throughput of the outputted tokens per second (TPS), and that of the parallel requests the hardware can handle (QPS). TPS is measured by running the model with a batch size of 1. It shows the pure latency overhead introduced by the TEE mode and reflects the performance of real-time requests. QPS is measured by maximizing the query throughput with a dynamically optimized batch size. It reflects the minimal average overhead the TEE mode brings.

We observed a difference in the impact of TEE mode between H100 and H200 in Table~\ref{tab:throughput_tee_on_ff}.
First, compared to H100, TEE mode introduces more overhead in H200 when runing the same model.
Second, the impact of TEE mode on QPS is more noticeable in H200.
In H100, the impact of TEE mode on TPS is nearly double that of QPS, but in H200, the difference decreased significantly.

\noindent\textbf{Conclusion 2: The overhead reduces as the model size grows.} As shown in Table \ref{tab:throughput_tee_on_ff}, the smallest model (\textbf{Llama-3.1-8B}) has the highest overhead. The medium-sized model (\textbf{Phi-3-14B-128k}) has roughly two-thirds of the overhead compared to the smaller one.
Notably, the largest model (\textbf{Llama-3.1-70B}) has a negligible overhead close to zero in H100.

\noindent\textbf{Conclusion 3: The latency is the main factor contributing to the overhead of the TEE mode.} Table \ref{tab:latency_tee_on_ff} shows the overhead introduced to the latency measured by TTFT and ITL. TTFT has a higher overhead compared with ITL, indicating the bottleneck is likely introduced by the I/O instead of the computation happening inside the TEE. Nevertheless, the overhead becomes trivial when hosting heavy computation models like \textbf{Llama-3.1-70B}.
Additionally, TEE mode has also a greater impact on TTFL and ITL in H200 compared to H100.

\begin{table}[htbp]
    \centering
    \begin{tabular}{llccc|ccc}
        \toprule
        \textbf{GPU}                   & \textbf{Model}         & \multicolumn{3}{c}{\textbf{TTFT (s)}} & \multicolumn{3}{c}{\textbf{ITL (s)}}                                                                                                                                                                                                \\
                                       &                        & \textbf{TEE-on}                       & \textbf{TEE-off}                     & \textbf{Overhead}                                                          & \textbf{TEE-on} & \textbf{TEE-off} & \textbf{Overhead}                                                          \\
        \midrule
        \multirow{3}{*}{\textbf{H100}} & \textbf{LLama-3.1-8B}  & 0.0288                                & 0.0242                               & 19.03\%                                                                    & 1.6743          & 1.5549           & 7.67\%                                                                     \\
                                       & \textbf{Phi3-14B-128k} & 0.0546                                & 0.0463                               & 18.02\%                                                                    & 3.7676          & 3.5784           & 5.29\%                                                                     \\
                                       & \textbf{Llama-3.1-70B} & 0.5108                                & 0.5129                               & -0.41\%\tablefootnote{The overhead is negative due to the precision loss.} & 94.8714         & 95.2395          & -0.39\%\tablefootnote{The overhead is negative due to the precision loss.} \\

        \midrule
        \multirow{3}{*}{\textbf{H200}} & \textbf{LLama-3.1-8B}  & 0.0364                                & 0.0301                               & 20.95\%                                                                    & 1.7158          & 1.5552           & 10.33\%                                                                    \\
                                       & \textbf{Phi3-14B-128k} & 0.0524                                & 0.0417                               & 25.60\%                                                                    & 3.6975          & 3.4599           & 6.87\%                                                                     \\
                                       & \textbf{Llama-3.1-70B} & 0.4362                                & 0.4204                               & 3.75\%                                                                     & 57.3855         & 55.9771          & 2.52\%                                                                     \\
        \bottomrule
    \end{tabular}
    \caption{Comparison of TTFT (Time to First Token) and ITL (Inter Output Token Latency) for TEE-on and TEE-off modes across models.}
    \label{tab:latency_tee_on_ff}
\end{table}

\begin{figure}
    \centering
    \begin{subfigure}[b]{0.49\linewidth}
        \centering
        \includegraphics[width=\linewidth]{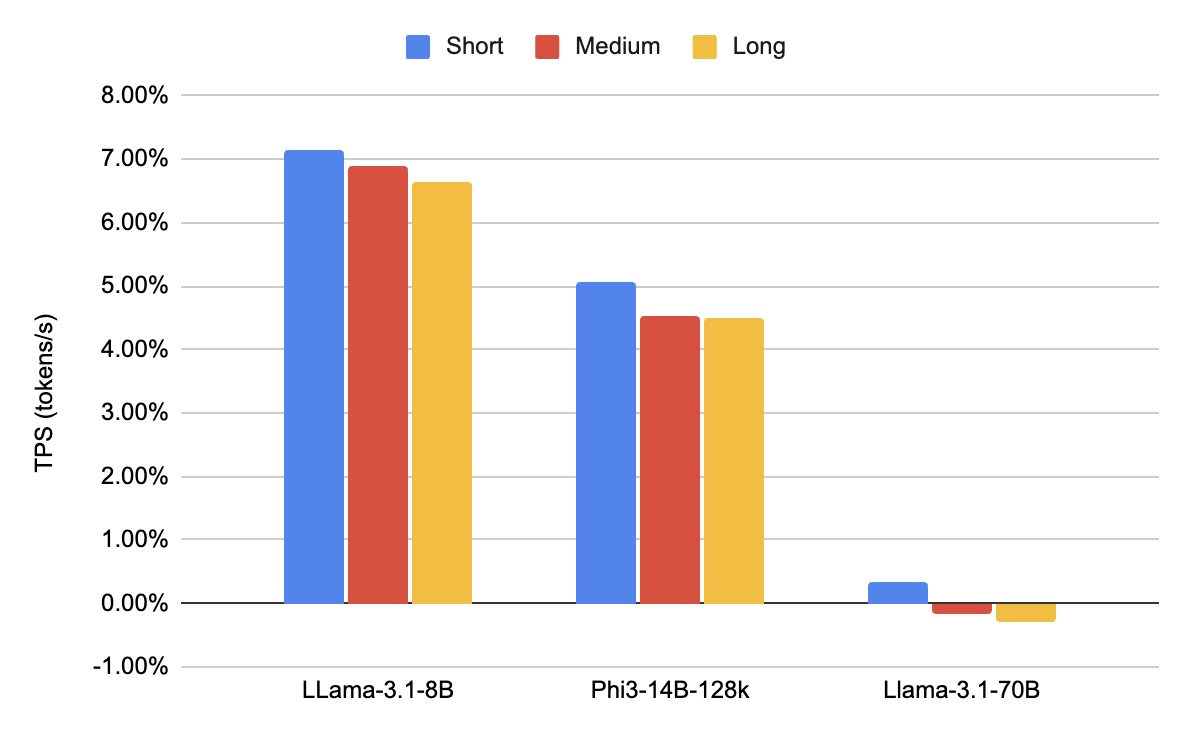}
        \caption{H100}
        \label{fig:tps-overhead-vs-token-size-h100}
    \end{subfigure}
    \begin{subfigure}[b]{0.49\linewidth}
        \centering
        \includegraphics[width=\linewidth]{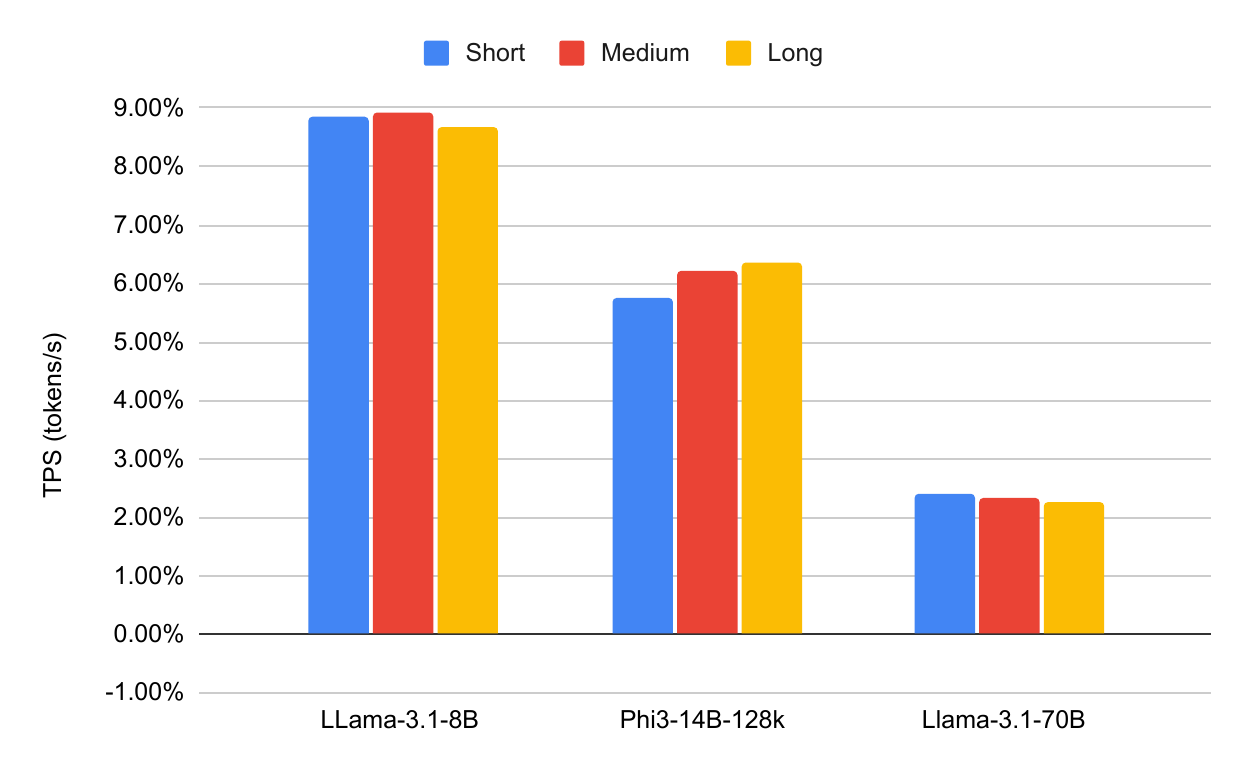}
        \caption{H200}
        \label{fig:tps-overhead-vs-token-size-h200}
    \end{subfigure}
    \caption{Throughput overhead across different token sizes (length of the input and output sequence). Short sequences are no longer than 100 tokens. Medium sequences are no longer than 500 tokens. Long sequences are between 501 and 1500 tokens.}
    \label{fig:tps-overhead-vs-token-size}
\end{figure}

\noindent\textbf{Conclusion 4: The overhead reduces as the token size grows.} As shown in Figure \ref{fig:tps-overhead-vs-token-size}, the throughput overhead reduces when the sequence length grows, measured by the total input and output token count. The detailed throughput metrics across various sequence lengths can be found in Table \ref{tab:tps_short_medium_long}.

\begin{table}[htbp]
    \centering
    \resizebox{\textwidth}{!}{%
        \begin{tabular}{llccc|ccc|ccc}
            \toprule
            \textbf{GPU}                   & \textbf{Model}         & \multicolumn{3}{c}{\textbf{TPS - short (tokens/s)}} & \multicolumn{3}{c}{\textbf{TPS - medium (tokens/s)}} & \multicolumn{3}{c}{\textbf{TPS - long (tokens/s)}}                                                                                                                                                                                                                                     \\
                                           &                        & \textbf{TEE-on}                                     & \textbf{TEE-off}                                     & \textbf{Overhead}                                  & \textbf{TEE-on} & \textbf{TEE-off} & \textbf{Overhead}                                                          & \textbf{TEE-on} & \textbf{TEE-off} & \textbf{Overhead}                                                          \\
            \midrule
            \multirow{3}{*}{\textbf{H100}} & \textbf{LLama-3.1-8B}  & 127.0310                                            & 136.8282                                             & 7.16\%                                             & 122.9356        & 132.0464         & 6.90\%                                                                     & 122.9705        & 131.7333         & 6.65\%                                                                     \\
                                           & \textbf{Phi3-14B-128k} & 70.9799                                             & 74.7556                                              & 5.05\%                                             & 66.1690         & 69.3104          & 4.53\%                                                                     & 66.2987         & 69.4176          & 4.49\%                                                                     \\
                                           & \textbf{Llama-3.1-70B} & 2.5983                                              & 2.6073                                               & 0.34\%                                             & 2.4413          & 2.4374           & -0.16\%\tablefootnote{The overhead is negative due to the precision loss.} & 2.5245          & 2.5168           & -0.30\%\tablefootnote{The overhead is negative due to the precision loss.} \\

            \midrule
            \multirow{3}{*}{\textbf{H200}} & \textbf{LLama-3.1-8B}  & 124.1744                                            & 136.2283                                             & 8.85\%                                             & 120.4250        & 132.2366         & 8.93\%                                                                     & 121.3849        & 132.9002         & 8.66\%                                                                     \\
                                           & \textbf{Phi3-14B-128k} & 71.8940                                             & 76.2754                                              & 5.74\%                                             & 67.8290         & 72.3372          & 6.23\%                                                                     & 68.5863         & 73.2384          & 6.35\%                                                                     \\
                                           & \textbf{Llama-3.1-70B} & 4.2261                                              & 4.3295                                               & 2.39\%                                             & 4.0425          & 4.1386           & 2.32\%                                                                     & 4.0947          & 4.1886           & 2.24\%                                                                     \\
            \bottomrule
        \end{tabular}%
    }
    \caption{Performance comparison of TEE-on and TEE-off modes across different sequence lengths in terms of TPS (tokens per second). Short sequences are no longer than 100 tokens. Medium sequences are no longer than 500 tokens. Long sequences are between 501 and 1500 tokens.}
    \label{tab:tps_short_medium_long}
\end{table}

\noindent\textbf{Conclusion 5: TEE can reach typical throughput.}
Here, we use NVIDIA H100 for the case study. Our experiments revealed that, with medium-sized inputs, the H100 GPU achieves 130 TPS for Llama-3.1-8B, while the larger Phi-3-14B model reaches approximately 6 TPS. These results demonstrate the robust performance of the H100 GPU across models of varying complexity.

\begin{figure}[ht!]
    \centering
    \includegraphics[width=1\linewidth]{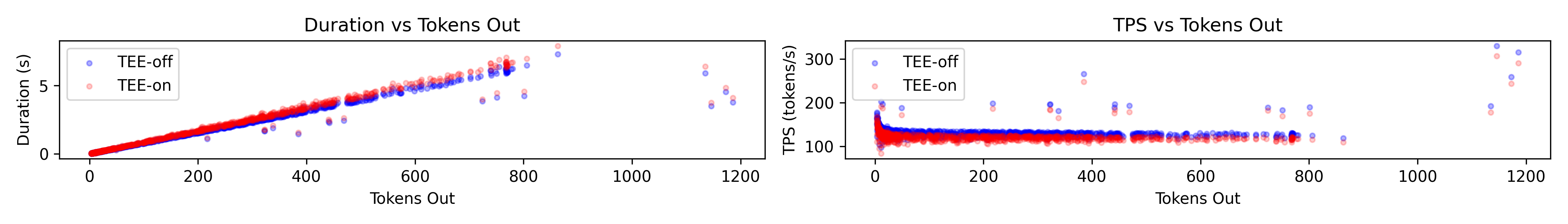}
    \caption{Throughput vs output token size for \textbf{LLama-3.1-8B} in H100.}
    \label{fig:scatter-llama8b-h100}
\end{figure}

\begin{figure}[ht!]
    \centering
    \includegraphics[width=1\linewidth]{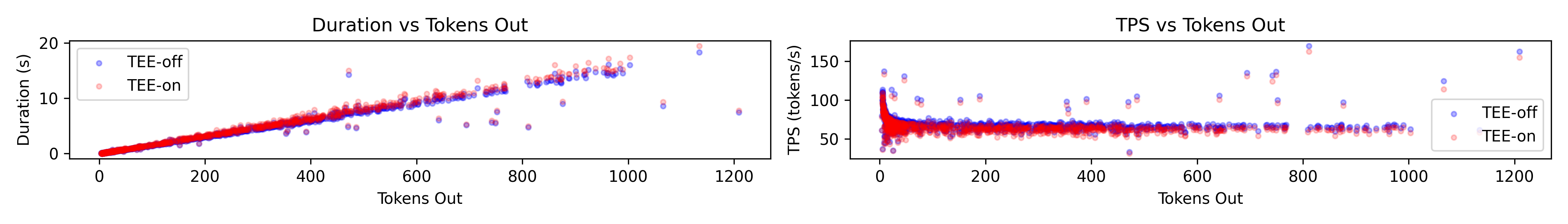}
    \caption{Throughput vs output token size for \textbf{Phi3-14B-128k} in H100.}
    \label{fig:scatter-phi14b-h100}
\end{figure}

\begin{figure}[ht!]
    \centering
    \includegraphics[width=1\linewidth]{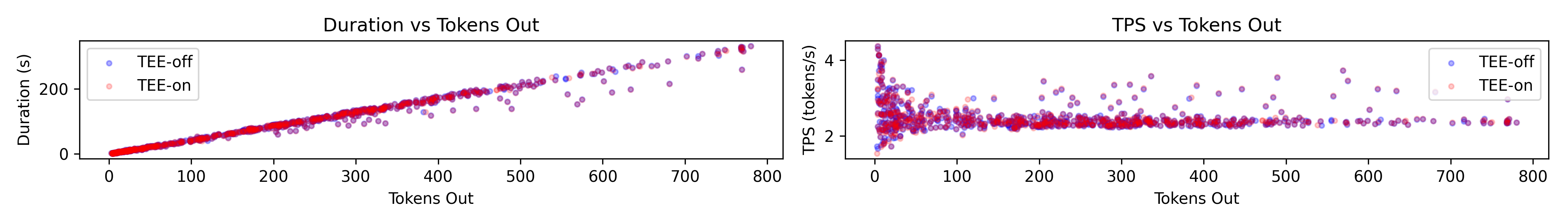}
    \caption{Throughput vs output token size for \textbf{Llama-3.1-70B} in H100.}
    \label{fig:scatter-llama70b-h100}
\end{figure}

\begin{figure}[ht!]
    \centering
    \includegraphics[width=1\linewidth]{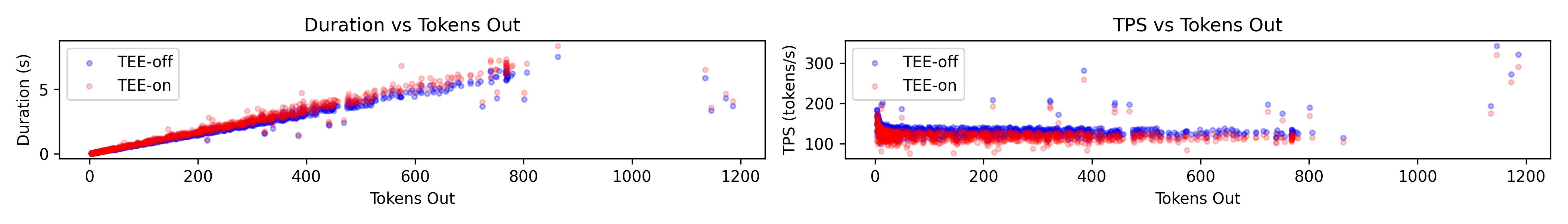}
    \caption{Throughput vs output token size for \textbf{LLama-3.1-8B} in H200.}
    \label{fig:scatter-llama8b-h200}
\end{figure}

\begin{figure}[ht!]
    \centering
    \includegraphics[width=1\linewidth]{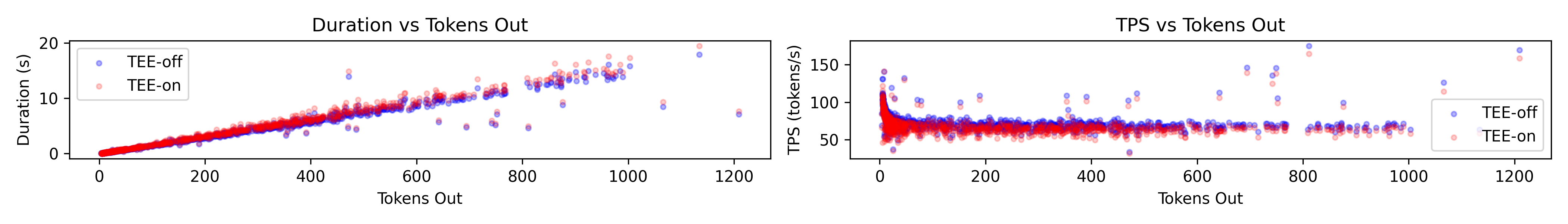}
    \caption{Throughput vs output token size for \textbf{Phi3-14B-128k} in H200.}
    \label{fig:scatter-phi14b-h200}
\end{figure}

\begin{figure}[ht!]
    \centering
    \includegraphics[width=1\linewidth]{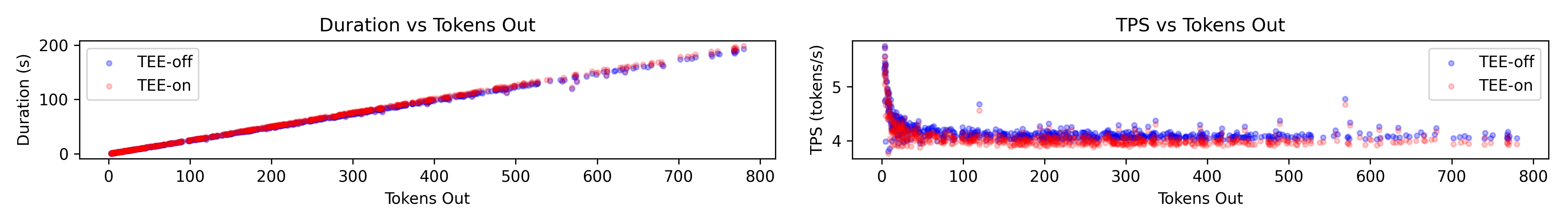}
    \caption{Throughput vs output token size for \textbf{Llama-3.1-70B} in H200.}
    \label{fig:scatter-llama70b-h200}
\end{figure}

More detailed experimental data for H100 is shown in Figures \ref{fig:scatter-llama8b-h100}, \ref{fig:scatter-phi14b-h100}, and \ref{fig:scatter-llama70b-h100}, and for H200 in Figures \ref{fig:scatter-llama8b-h200}, \ref{fig:scatter-phi14b-h200}, and \ref{fig:scatter-llama70b-h200}.

\section{Conclusion}

Our results show that as input size grows, the efficiency of TEE mode increases significantly. When computation time within the GPU dominates overall processing time, the I/O overhead introduced by TEE mode diminishes, allowing efficiency to approach nearly 99\%.

Efficiency growth is more pronounced in larger models, such as \textbf{Phi3-14B-128k} and \textbf{Llama-3.1-70B}, due to their greater computational demands, which result in longer GPU processing times. Consequently, the I/O overhead becomes increasingly trivial as model size increases.

The total token size (sum of input and output token size) significantly influences the throughput overhead. Larger total token counts lead to higher efficiencies, as they enhance the ratio of computation time to I/O time.

These findings underscore the scalability of TEE mode in handling large-scale LLM inference tasks, particularly as input sizes and model complexities grow. The minimal overhead in high-computation scenarios validates its applicability in secure, high-performance AI workloads.

\bibliographystyle{alpha}
\bibliography{references}

\end{document}